\documentclass[reprint,superscriptaddress,10pt, amsmath,amssymb, aps]{revtex4-2}
\pdfoutput=1 
\usepackage[utf8]{inputenc}
\usepackage[T1]{fontenc}
\usepackage{graphicx}             
\usepackage{xcolor}
\definecolor{darkblue}{rgb}{0, 0, 0.8}
\usepackage[colorlinks=true, breaklinks=true, linkcolor=darkblue, citecolor=darkblue, urlcolor=darkblue]{hyperref}
\usepackage{braket}
\usepackage{soul} 

\newcommand*{\tx}[1]{\mathrm{#1}} 
\newcommand*{\us}{\,}             

\definecolor{green2}{rgb}{0., 0.5, 0.}
\definecolor{blue2}{rgb}{0., 0.35, 1}
\definecolor{brown}{rgb}{0.7, 0.15, 0}
\definecolor{red}{rgb}{1, 0.1, 0.1}

\begin{document}

\title{Accelerating Polaritons with External Electric and Magnetic Fields}

\author{T. Chervy}
\thanks{These authors contributed equally: T. Chervy and P. Kn\"uppel.}
\affiliation{Institute of Quantum Electroncis, ETH Z\"urich, CH-8093 Z\"urich, Switzerland}

\author{P. Kn\"uppel}
\thanks{These authors contributed equally: T. Chervy and P. Kn\"uppel.}
\affiliation{Institute of Quantum Electroncis, ETH Z\"urich, CH-8093 Z\"urich, Switzerland}

\author{H. Abbaspour}
\affiliation{Institute of Quantum Electroncis, ETH Z\"urich, CH-8093 Z\"urich, Switzerland}

\author{M. Lupatini} 
\affiliation{Solid State Physics Laboratory, ETH Z\"urich, CH-8093 Z\"urich, Switzerland}

\author{S. F\"alt}
\affiliation{Institute of Quantum Electroncis, ETH Z\"urich, CH-8093 Z\"urich, Switzerland}
\affiliation{Solid State Physics Laboratory, ETH Z\"urich, CH-8093 Z\"urich, Switzerland}

\author{W. Wegscheider}
\affiliation{Solid State Physics Laboratory, ETH Z\"urich, CH-8093 Z\"urich, Switzerland}

\author{M. Kroner}
\affiliation{Institute of Quantum Electroncis, ETH Z\"urich, CH-8093 Z\"urich, Switzerland}

\author{A. Imamo\v{g}lu}
\affiliation{Institute of Quantum Electroncis, ETH Z\"urich, CH-8093 Z\"urich, Switzerland}

\date{\today}

\begin{abstract}
It is widely assumed that photons cannot be manipulated using electric or magnetic fields. Even though hybridization of photons with electronic polarization to form exciton-polaritons has paved the way to a number of ground-breaking experiments in semiconductor microcavities, the neutral bosonic nature of these quasiparticles has severely limited their response to external gauge fields. Here, we demonstrate polariton acceleration by external electric and magnetic fields in the presence of nonperturbative coupling between polaritons and itinerant electrons, leading to formation of new quasiparticles termed polaron-polaritons. We identify the generation of electron density gradients by the applied fields to be primarily responsible for inducing a gradient in polariton energy, which in turn leads to acceleration along a direction determined by the applied fields. Remarkably, we also observe that different polarization components of the polaritons can be accelerated in opposite directions when the electrons are in $\nu = 1$ integer quantum Hall state.
\end{abstract}
\maketitle

Controlling photons with external electric or magnetic fields is an outstanding goal. On the one hand, coupling photons to artificial gauge fields holds promises for the realization of topological and strongly correlated phases of light \cite{lu_topological_2014, ozawa_topological_2019, aidelsburger_artificial_2018, carusotto_quantum_2013}. On the other hand, effecting forces on photons constitutes both a problem of fundamental interest in electromagnetism and an important step in view of technological applications \cite{obrien_photonic_2009, sanvitto_road_2016, schneider_exciton-polariton_2016, engheta_circuits_2007}.
One promising avenue towards this goal is to hybridize photons with material excitations that are genuinely sensitive to gauge fields \cite{lim_electrically_2017}. In this non-perturbative regime, exciton-polariton states are formed, ensuring that the forces acting on the material excitations are directly imprinted onto the photon. However, the neutral bosonic nature of polaritons has so far severely limited their response to gauge fields \cite{karzig_topological_2015, bleu_measuring_2018, gutierrez-rubio_polariton_2018, klembt_exciton-polariton_2018}.

A particularly appealing approach to circumvent this limitation is to leverage on the interaction between excitons and charges. Indeed, early reports on the drift of trions in an electric field \cite{sanvitto_observation_2001, pulizzi_optical_2003}, as well as on the Coulomb drag effect in bilayer systems \cite{kulakovskii_screening_2004, berman_can_2010, berman_drag_2010, narozhny_coulomb_2016} indicated that it may be possible to manipulate neutral excitations using electric fields in a solid-state setting. Recently, experimental \cite{myers_pushing_2018} and theoretical studies \cite{chestnov_pseudo-drag_2019} reported the electrical control of the speed of a polariton superfluid, raising new questions and possibilities regarding the interplay between the normal and condensed fractions of the fluid in the presence of electron-exciton interactions.

While interactions between polaritons and electrons have been proposed and analyzed as a mechanism for polariton thermalization \cite{malpuech_polariton_2002, lagoudakis_electron-polariton_2003, qarry_nonlinear_2003, perrin_polariton_2005}, early studies reported the modifications to polariton resonances in the presence of a Fermi sea \cite{brunhes_oscillator_1999, rapaport_negatively_2000, rapaport_negatively_2001}. These modifications stem form dispersive interactions between the polarizable excitonic component of the polariton with the charge-density fluctuations of the Fermi sea \cite{suris_correlation_2003, combescot_trion_2003,combescot_many-body_2005}. In a typical setting where excitons are hosted by a GaAs quantum well (QW), the existence of a spin-singlet trion bound channel leads to the formation of two optically active branches referred to as attractive and repulsive polarons~\cite{sidler_fermi_2017,efimkin_many-body_2017}. These polaronic states consist of an exciton dressed by collective trion-hole excitations of the Fermi sea, where the spin of the photo-excited electron is opposite to that of the dressing Fermi sea electrons \cite{sidler_fermi_2017, efimkin_many-body_2017, efimkin_exciton-polarons_2018}. Upon decreasing the electron density, the quasi-particle weight of the repulsive polaron branch, quantifying its excitonic character, increases. In the limit of vanishing electron density, the repulsive (attractive) polaron asymptotically becomes the bare exciton (molecular trion) state with strong (vanishing) coupling to the cavity mode \cite{sidler_fermi_2017}. The many-body polaronic states are expected to be charge neutral \cite{combescot_many-body_2005,sidler_fermi_2017}, suggesting the absence of coupling to external fields. In a recent theoretical study however, it was shown that neutral polarons are sensitive to the average force on electrons, leading to a finite polaron trans-conductivity in the non-equilibrium limit -- an effect that is observable even when polarons hybridize with cavity photons \cite{cotlet_transport_2019}. 

In the present work, we demonstrate experimentally that external electric and magnetic fields effect forces on polaron-polaritons. In contrast to earlier proposals, we find that the observed polariton acceleration primarily originates from a source-drain voltage induced density gradient in the two dimensional electron gas (2DEG) in which polaritons are immersed. Remarkably, we show that the direction of the resulting force can be changed by an externally applied magnetic field, since the induced Hall voltage creates transverse density gradients. Finally, we extend this approach to demonstrate spin-selective acceleration of polaritons when the 2DEG is in the integer quantum Hall regime.

\begin{figure}
\includegraphics{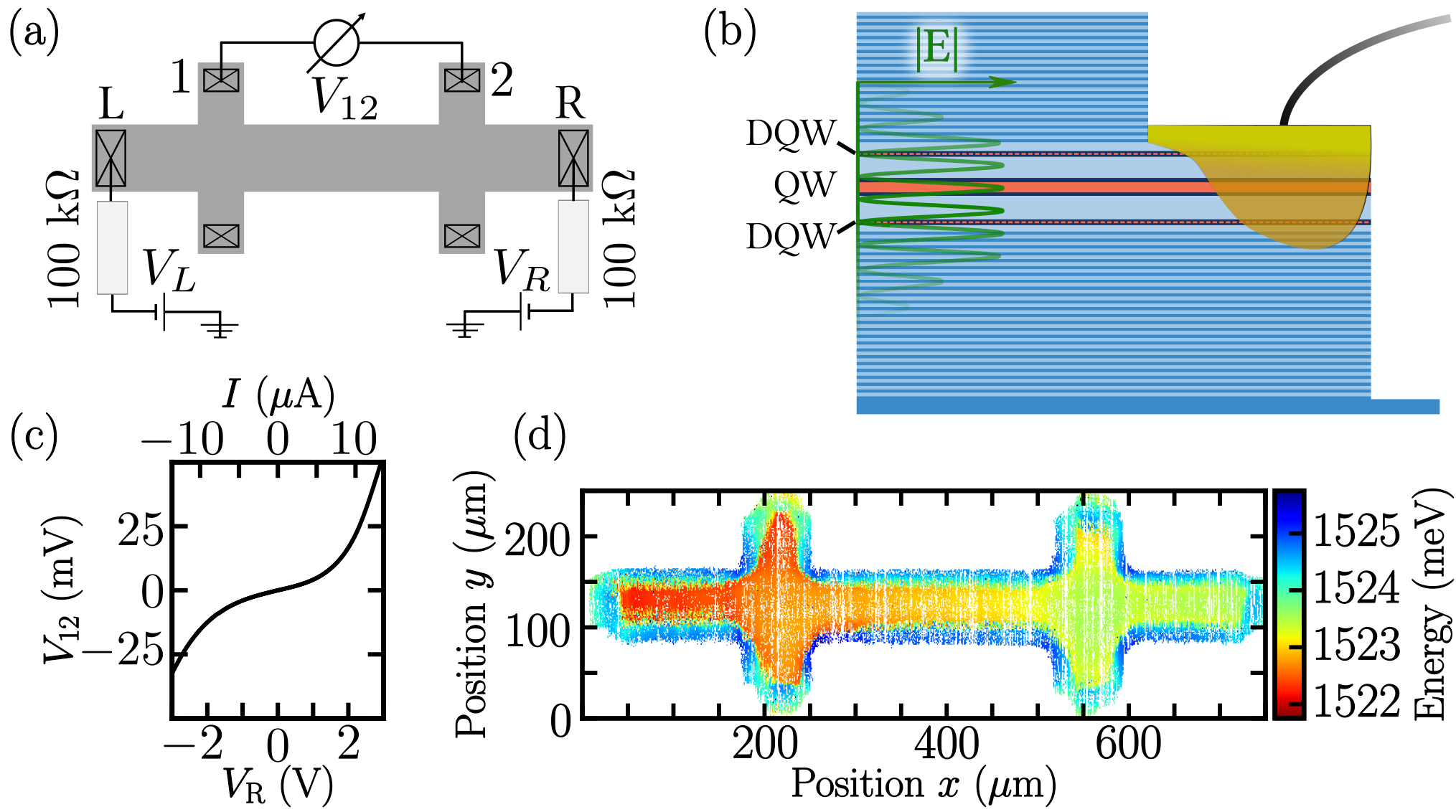}
\caption{\label{fig:Figure_1} Electrical and optical properties of the sample. (a) Layout of the Hall bar and the contacts. (b) Side-view of the Hall bar. The top mirror has been etched to contact the 2DEG located in the center QW. DQW, doping quantum wells. (c) Current-voltage characteristic of the device where $V_\tx{R}$ acts as current source and $V_\tx{L}$ as drain. (d) Map of the lower polariton PL wavelength measured at normal incidence ($k_\parallel = 0$).}
\end{figure}

The sample used in this study is a monolithic distributed Bragg reflector optical microcavity with a quality factor $Q = 5.5 \cdot 10^3$ grown by molecular-beam epitaxy. The structure contains a single optically active GaAs QW of $20 \us \mathrm{nm}$ thickness located at the central antinode of the cavity field. This QW is remotely doped from both sides by Si impurities embedded in thin GaAs QWs ($10 \us \mathrm{nm}$ thickness) located at the nodes of the cavity field. The doping QWs thus provide electrons to the optically active QW where a 2DEG is formed with nominal electron density of $n_e = 0.33 \cdot 10^{11} \us \tx{cm}^{-2}$ and mobility $\mu = 1.6 \cdot 10^{6} \us \tx{cm}^{2} \us \tx{V}^{-1} \us \tx{s}^{-1}$ (see \cite{ravets_polaron_2018} for further details on the sample fabrication). To study the interplay between polariton transport and 2DEG physics, the sample is etched in the form of a Hall bar with annealed electrical contacts to the 2DEG, as depicted in Fig. 1(a,b). A photo-luminescence (PL) scan map of the sample, recorded at normal incidence (in-plane momentum $k_\parallel = 0$), is shown in Fig. 1(d) outlining the shape of the Hall bar. The increase in emission energy of about $3 \us \tx{meV/mm}$ from left to right is due to a wedge in the cavity length introduced during sample growth.
The sample is cooled to a base temperature of $20 \us \tx{mK}$ in a dilution refrigerator with free space optical access enabling simultaneous position and momentum resolved microscopy, as well as magneto-transport experiments (see Supplemental Material \cite{noauthor_see_nodate}). 

We first characterize the electrical properties of our sample by a four-point current-voltage (I-V) measurement yielding the characteristic shown in Fig. 1(c). At low source-drain voltage bias the I-V curve is linear with a 2DEG resistivity of $210 \us \tx{\Omega/sq}$. Increasing the source-drain bias to ca. $\pm 1 \us \tx{V}$, the I-V curve becomes nonlinear, entering a regime where the externally applied electro-chemical potential leads to a spatially varying chemical potential of the 2DEG. At even larger source-drain bias, the 2DEG is depleted and we recover a linear I-V characteristic with an increased resistivity of $1500 \us \tx{\Omega/sq}$, corresponding to electrical conduction in parallel channels (see Supplemental Material \cite{noauthor_see_nodate}).

\begin{figure}
\includegraphics{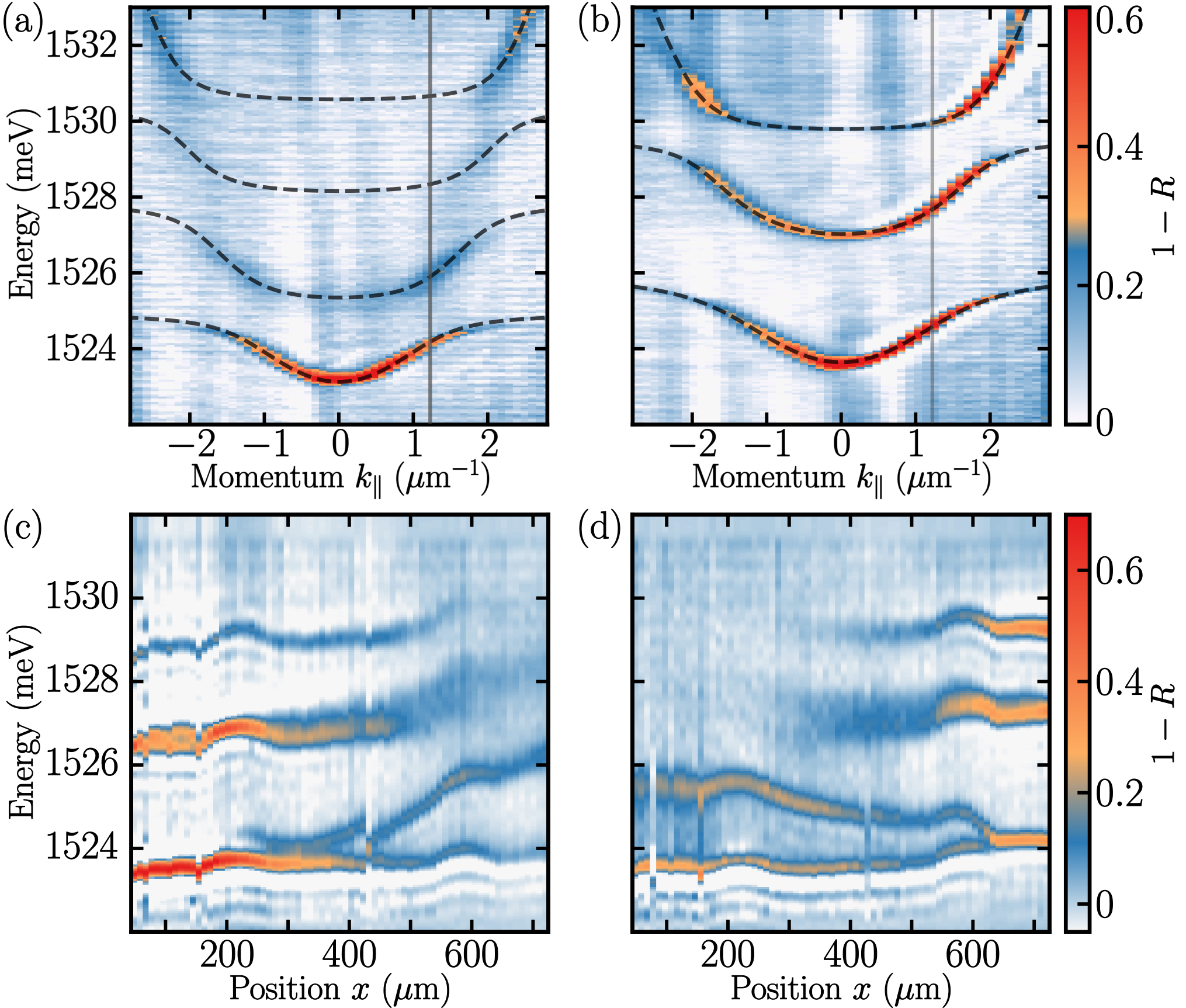}
\caption{\label{fig:Figure_2} 
Electrically controlled polariton landscape. (a) Normalized white-light reflectivity spectra showing the polaron-polariton dispersion at the nominal electron density. (b) Exciton-polariton dispersion in the depleted regime ($\Delta V_\tx{L} = -6 \us \tx{V}$). Dashed lines are coupled oscillators fits to the data. (c,d) Normalized white-light reflectivity spectra at $k_{\parallel} = 1.2 \us \tx{\mu m}^{-1}$ (vertical line in (a,b)) as a function of position with a negative bias voltage applied to the left contact $(\Delta V_\tx{L} = -2.4 \us \tx{V})$ (c) and to the right contact $(\Delta V_\tx{R} = -2.4 \us \tx{V})$ (d). }
\end{figure}

These different regimes of the 2DEG behavior have a counterpart on the optical response of the system. Figure 2(a) shows the angular dispersion of the polaron-polariton states at zero source-drain bias. There, we identify four branches resulting from the hybridization of the cavity photon with the attractive and repulsive heavy-hole polarons and with the light-hole resonance. The dispersive polaron-polariton branches are well reproduced by the coupled oscillators fit shown in dashed lines. Figure 2(b) shows the angular dispersion acquired while applying a large bias of $-6 \us \tx{V}$ to the left contact and putting the right contact to ground, $\Delta V_\tx{L} = -6V$ (see Fig. 1(a)). There, we measure the typical exciton-polariton dispersion of an intrinsic QW with anti-crossings about the heavy-hole and light-hole exciton energies, as expected from the 2DEG depletion observed in the I-V characteristic.

In between these two extremes, at intermediate bias voltages, the electron density shows smooth spatial gradients across the Hall bar as demonstrated in Fig. 2(c) for $\Delta V_\tx{L} = -2.4 \us \tx{V}$. Here, the polariton spectrum is recorded at positions across the long axis of the Hall bar and at fixed collection angle corresponding to $k_{\parallel} = 1.2 \us \tx{\mu m}^{-1}$. On the left side, where the negative bias is applied, the sample is devoid of electrons and the spectrum resembles a vertical cut in the dispersion of Fig. 2(b). Moving to the right, the electron density increases and we gradually recover polaron-polariton spectra corresponding to the dispersion of Fig. 2(a) as the oscillator strength is gradually transferred from the excitonic to the polaronic resonances. Reversing the applied electric field ($\Delta V_\tx{R} = -2.4 \us \tx{V}$) inverts this density gradient, as shown in Fig. 2(d). Such electron density gradients constitute electrically tunable potential landscapes for polaritons which we explore in the following to transport neutral optical excitations.

\begin{figure}
\includegraphics{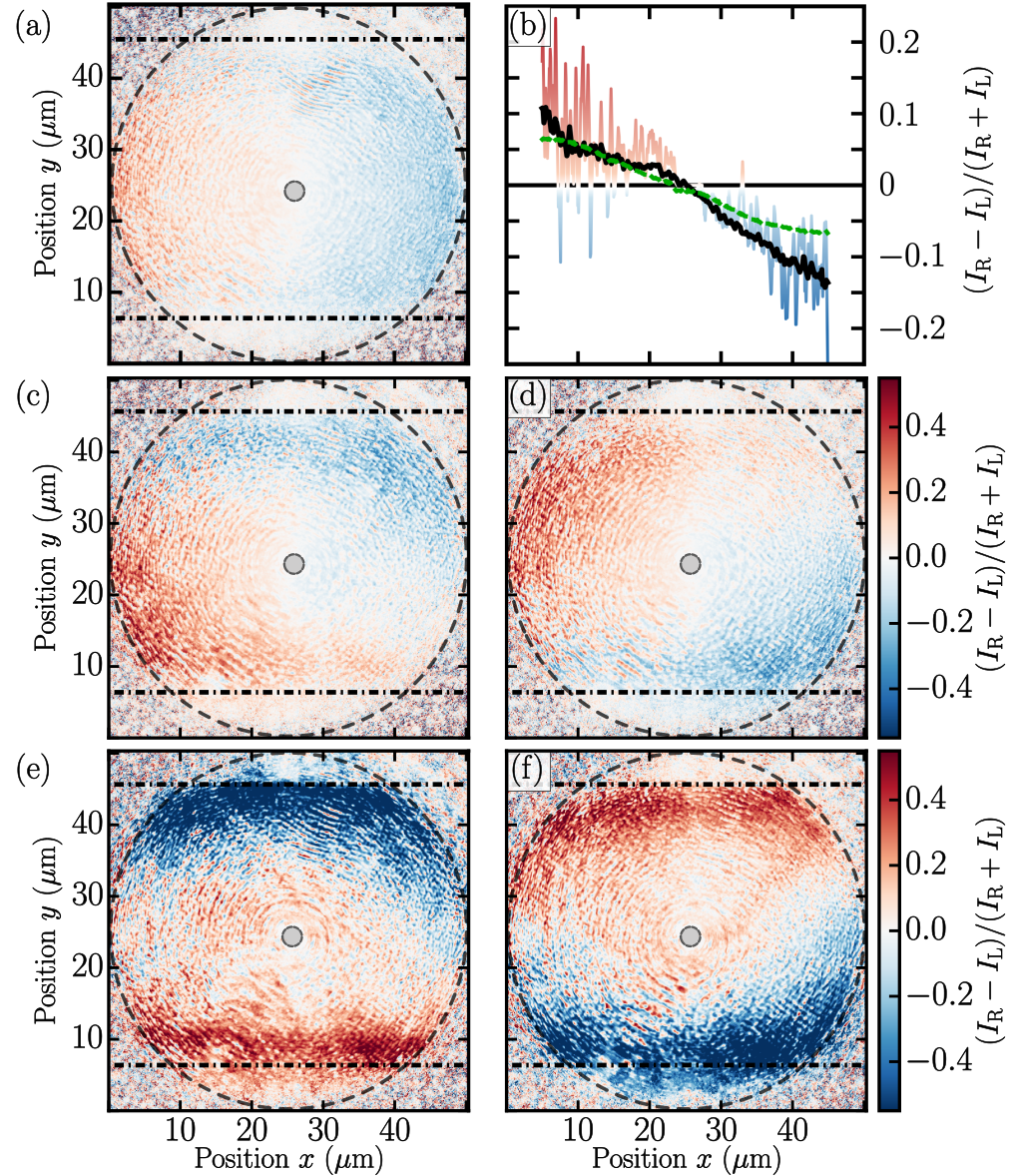}
\caption{\label{fig:Figure_3} Polariton acceleration in external electric and magnetic fields. (a) Normalized difference between two RF images of polariton flow in opposing electron density gradients. (b) Comparison with a trajectory based model (dashed green line). The red-blue line is a linecut of panel (a) at $y=25 \us \tx{\mu m}$ and the black line is an average between $y=10$ and $y=40 \us \tx{\mu m}$. (c-f) Normalized difference between images $I_\tx{R}$, $I_\tx{L}$ as in (a) for Magnetic fields: (c) $B=8 \us \tx{mT}$, (d) $B=-8 \us \tx{mT}$ (e), $B=40 \us \tx{mT}$ and (f) $B=-40 \us \tx{mT}$. The horizontal dashed-dotted lines in (a, c-f) delimit the width of the Hall bar, the large circle is the field of view of the microscope and the central grey disk is the excitation spot.}
\end{figure}

To demonstrate polariton transport, we resonantly excite a polariton cloud and image its expansion in two opposing electron density gradients. The excitation laser, at $1524.0 \us \tx{meV}$ $(813.54 \us \tx{nm})$, is linearly polarized and focused in the central region of the Hall bar ($x=450 \us \tx{\mu m}$ in Fig. 1 and 2), thus injecting a radially expanding cloud of polaritons with finite $k_\parallel$. The decaying polariton signal is collected by the same microscope objective lens and separated from the scattered excitation beam by polarization filtering. A finite strain along the crystalline axes in the structure allows us to obtain this resonance fluorescence (RF) signal by polarizing the excitation beam at $45^\circ$ with respect to the polarization eigenbasis defined by strain (see Supp. \cite{noauthor_see_nodate}). Two images, $I_\tx{R}$ and $I_\tx{L}$, of this RF signal are acquired under source-drain biases of $\Delta V_\tx{R}= -2.4 \us \tx{V}$ and $\Delta V_\tx{L} = -2.4 \us \tx{V}$, respectively. These two voltages were chosen such that at the injection spot on the Hall bar, the electron densities are the same and the gradients are of opposite signs. This choice is not necessary but simplifies the observation of polariton acceleration as there is no trivial difference between dispersions and thereby group velocities between the two images to be compared. Figure 3(a) shows the normalized difference of the two images $(I_{\tx{R}} - I_{\tx{L}}) / (I_{\tx{R}} + I_{\tx{L}})$ clearly demonstrating the ability to route polaritons by electrical means.   

In order to model this effect, we measure the polariton dispersion at the two sides of the field of view ($x=0$ and $x=50 \us \tx{\mu m}$ in Fig. 3(a)) and perform a coupled oscillator fit. By linearly interpolating across the field of view, we characterize the lower polariton energy landscape $E_\tx{LP}(x, k_x)$, separately for the two density gradients used to measure $I_\tx{R}$ and $I_\tx{L}$ (see Supplementary Material \cite{noauthor_see_nodate}). Starting from the classical Hamilton equations of motion in one dimension ($x$), we propagate the experimentally determined initial conditions $k_{x}(0) = \pm 1.2 \us \tx{\mu m}^{-1}$ to predict the emission intensity $I(x)$, again repeated for the two energy landscapes corresponding to the two opposing gradients. As in the experiment, we calculate the normalized difference between the two cases and compare the result (green dashed line) with the experiment in Fig. 3(b). The red-blue colored line is a line cut through Fig. 3(a) at $y=25 \us \tx{\mu m}$ and the black curve is an average over $y$ between $10$ and $40 \us \tx{\mu m}$. Remarkably, this simple approach allows us to obtain reasonable quantitative agreement with our data.

We can further control the polariton flow by applying a magnetic field perpendicular to the QW plane. In conjunction with a finite source-drain bias, this induces a Hall voltage transverse to the applied potential leading to charge redistribution in the $y$-direction \cite{fontein_spatial_1991, cage_potential_1995}. The combined electric and magnetic fields now shape the electron density gradient which can be tuned in angle and magnitude, as demonstrated in Fig. 3(c-f). Figure 3(c) corresponds to alternating voltage biases of $\Delta V_\tx{R}= -2.4 \us \tx{V}$ and $\Delta V_\tx{L} = -2.4 \us \tx{V}$ and a fixed magnetic field of $8 \us \tx{mT}$ and shows the possibility to orient the polariton flow in a diagonal direction. Further increasing the magnetic field to $40 \us \tx{mT}$ (Fig. 3(e)) leads to polariton transport in the up-down direction with an intensity contrast of \textit{ca.} $50\%$. Moreover, flipping  the sign of the applied magnetic field reverses the direction of polariton transport, as shown in Fig. 3(d,f).

While these results demonstrate the possibility to transport dressed photons (\textit{i.e.} polaritons) by electric and magnetic fields, it should be noted that we do not observe here a Lorentz force for photons. In particular, the force acting on polaritons does not appear to depend on the direction of their motion. As clearly shown in Fig. 3(e,f), both the polaritons propagating to the left and to the right are deflected in the same direction. The polariton acceleration is determined only by the electron density gradient which in turn is controlled by the combination of magnetic field and electrical bias. In other words, the nonperturbative coupling of polaritons to itinerant electrons allows for the control of photons by electromagnetic forces acting on the electronic sector. In the last part of this article, we demonstrate how this idea can be extended to realize transverse polariton spin currents reminiscent of an intrinsic spin-Hall effect when the 2DEG is close to the $\nu=1$ integer quantum Hall state.

\begin{figure}
\includegraphics{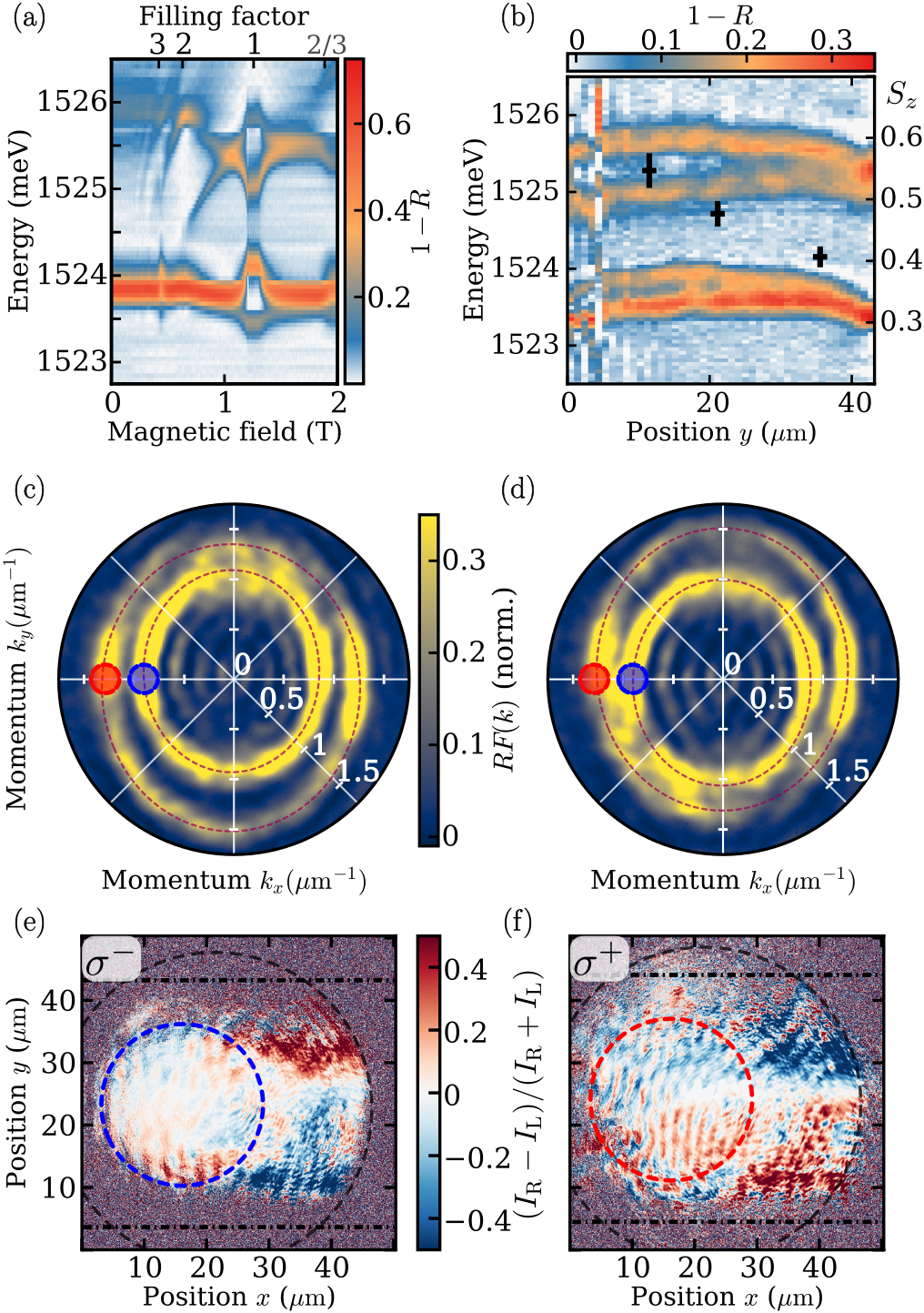}
\caption{\label{fig:Figure_4} Spin-selective polariton acceleration. (a) Normalized white light reflectivity spectra measured at $k_\parallel = 1.2 \us \tx{\mu m}^{-1}$ as a function of magnetic field. The corresponding Landau level filling factor is determined from an independent magneto-transport measurement. (b) Normalized reflectivity spectra at $k_\parallel = 1.2 \us \tx{\mu m}^{-1}$ and $B = 1.1 \us \tx{T}$ across the vertical $y$-direction on the Hall bar, with a source-drain bias $\Delta V_\tx{R}= -0.17 \us \tx{V}$. Black points indicates the electron spin polarization $S_z$ measured at 3 different positions. (c,d) Momentum-resolved polariton RF emission under cross-linear polarization at $1523.8 \us \tx{meV}$  $(813.62 \us \tx{nm})$ for (c) $\Delta V_\tx{R}= -0.17 \us \tx{V}$ and (d) $\Delta V_\tx{L}= -0.35 \us \tx{V}$. The red dashed lines are guides to the eye. (e,f) Normalized difference between two RF images of right-propagating polaritons acquired with $\Delta V_\tx{R}= -0.17 \us \tx{V}$ and $\Delta V_\tx{L}= -0.35 \us \tx{V}$. (e) Excitation of $\sigma^-$ polaritons at $k_x = -0.9 \us \tx{\mu m}^{-1}$, $k_y=0$. (f) Excitation of $\sigma^+$ polaritons at $k_x = -1.3 \us \tx{\mu m}^{-1}$, $k_y=0$. The horizontal dashed-dotted lines delimit the width of the Hall bar, the large circle is the field of view of the microscope. The red and blue rings in (c-f) show the excitation spots in real and momentum spaces for $\sigma^+$ and $\sigma^-$ polaritons, respectively.}
\end{figure}

At low temperature and under a strong magnetic field, the excitation spectrum of a high mobility 2DEG exhibits energy gaps due to the quantization of cyclotron orbits. These Landau levels are further split by a Zeeman field, forming a ladder of spin sub-bands for the electrons. As the occupancy of this ladder is varied (\textit{e.g.} by tuning the magnetic field), the 2DEG undergoes phase transitions between ground states of different spin polarization $S_z$. The interplay between the spin polarization of the 2DEG and the spin of the optically generated electron leads to dramatic modifications of the spin-singlet polaron-polariton spectrum, as shown in Fig. 4(a) and discussed in details in Refs. \cite{smolka_cavity_2014, ravets_polaron_2018}. In particular, at filling factor $\nu = 1$ ($1.26 \us \tx{T}$), we observe a strong reduction in the attractive polaron-polariton Rabi splitting $\Omega$ for left-hand circularly polarized light $\sigma^-$ and a concomitant increase for right-hand circularly polarized light $\sigma^+$, revealing the spin polarization of the 2DEG at this integer quantum Hall plateau $S_z\simeq(\Omega_{\sigma^+}^2-\Omega_{\sigma^-}^2)/(\Omega_{\sigma^+}^2+\Omega_{\sigma^-}^2)\simeq 70\%$ \cite{ravets_polaron_2018}. In this quantum Hall regime, the charge density gradients demonstrated above translate into gradients of 2DEG spin polarization, resulting in optical spin-contrasted forces for polaritons.

Figure 4(b) shows the evolution of the polariton spectrum at $k_{\parallel} = 1.2 \us \tx{\mu m}^{-1}$, recorded from the lower edge to the upper edge of the Hall bar ($y$-direction) under a voltage bias of $\Delta V_\tx{R}= -0.17 \us \tx{V}$ and a magnetic field of $1.1 \us \tx{T}$. We observe four energy branches, as expected from a vertical cut in Fig. 4(a) near $\nu=1$, where the two inner branches are the $\sigma^-$-polarized lower and upper polaritons and the two outer ones are the $\sigma^+$-polarized lower and upper polaritons. As can be seen from this panel, the degree of electron spin polarization of the 2DEG $S_z$ evolves across the Hall bar and can be controlled electrically as shown in Supplemental Material \cite{noauthor_see_nodate}. Remarkably, such variations in $S_z$ constitute gradients of opposite signs for the $\sigma^+$ and $\sigma^-$ polaron-polariton energy landscapes. 

To investigate the resulting spin-dependent polariton acceleration, we first perform momentum-resolved measurements by imaging the polariton RF emission (see Supplemental Materials \cite{noauthor_see_nodate}). An excitation energy of $1523.8 \us \tx{meV}$  $(813.62 \us \tx{nm})$ is chosen to intercept both $\sigma^+$ and $\sigma^-$ lower polariton dispersions at finite $k_{\parallel}$. Figure 4(c,d) show the momentum-resolved polariton RF signal for $\Delta V_\tx{R}= -0.17 \us \tx{V}$ and $\Delta V_\tx{L}= -0.35 \us \tx{V}$ respectively. As can be seen in Fig. 4(c), the inner $\sigma^-$-polarized branch is shifted towards positive $k_y$ while the outer $\sigma^+$-polarized branch is shifted towards negative $k_y$. This observation directly demonstrates an in-plane acceleration whose sign depends on the polaritons spin polarization in the given quantum state $S_z^\tx{pol}$, that is the component of the polariton pseudo-spin normal to the sample surface \cite{kavokin_quantum_2004}. The effect is reversed in Fig. 4(d), where the external bias and thereby the gradient in electron spin polarization points in the opposite $y$-direction. The real-space counterpart of this acceleration allows for the generation of transverse optical spin currents, reminiscent of an intrinsic optical spin-Hall effect \cite{onoda_hall_2004, kavokin_optical_2005, leyder_observation_2007, amo_anisotropic_2009, maragkou_optical_2011, lekenta_tunable_2018, gianfrate_direct_2019}. 

To demonstrate the generation of transverse polariton spin currents, we inject polariton waves of well defined momenta by focusing the excitation beam in the back-focal plane of the objective lens. Figure 4(e,f) correspond to excitation at $k_x = -0.9 \us \tx{\mu m}^{-1}$, $k_y=0$ and $k_x = -1.3 \us \tx{\mu m}^{-1}$, $k_y=0$, resulting in right propagating polaritons with $\sigma^-$ and $\sigma^+$ polarization, respectively. The normalized difference of the propagation images, acquired with the two different voltage biases, indeed reveals opposite acceleration of $\sigma^-$ and $\sigma^+$ polaritons along the $y$-direction.

The spin-dependent momentum shifts demonstrated here, although capable of generating transverse spin currents, remains fundamentally different from the usual Rashba type coupling at the origin of standard spin-Hall effects. Instead, the interaction reported here is analogous to a force for photons, where the spatially varying electron spin polarization ($S_z$) acts as an accelerating potential, sorting polaritons of different spin ($S_z^\tx{pol}$) in different directions: 
\begin{equation}
        {\bf F}_{\tx{Photon}} \sim S_z^\tx{pol} {\bf\nabla} S_z.
\end{equation}
It should be noted that the evolution of electron spin polarization around $\nu=1$ quantum Hall plateau is widely believed to involve the proliferation of skyrmions in the quantum Hall ferromagnetic state due to the interplay between Zeeman and Coulomb energies \cite{girvin_spin_2015}. The spin-singlet polaron-polariton dressing thus constitutes a new interface for coupling the optical spin of photons to the electronic spin excitations of 2DEGs. The behavior of such interactions in the fractional quantum Hall regime where excitons may be dressed by fractionally charged quasi-particles, as well as the residual interactions between these fractional quantum Hall polaritons remain to be explored \cite{knuppel_nonlinear_2019}.

In summary, we demonstrated novel ways to control polariton flows using external electric and magnetic fields. A non-equilibrium electron density gradient acts as an effective electric field for polaron-polaritons and is tunable in strength and direction. We foresee that this effective electric field could be further controlled by tailoring the 2DEG density \textit{e.g.} using patterned electrodes. By mapping the energy landscape of the lower polariton, we reach quantitative agreement between a simple trajectory based model and the observed polariton acceleration. Our experiment constitutes an alternative to the already proposed polariton drag effect for effecting electro-magnetic forces on neutral optical excitations \cite{berman_drag_2010, cotlet_transport_2019}. We emphasize that the electron density gradients we exploit are generic for low density 2DEGs when large source-drain voltages are applied and therefore need to be considered in view of polariton drag experiments. In the integer quantum Hall regime, we demonstrate that electron spin depolarization, induced by the proliferation of skyrmions as the electron density gradient pushes the system away from $\nu =1$ filling, constitutes a scalar potential for the optical spin of photons and results in spin-dependent accelerations. This observation suggests that magnetic excitations of a spin-ordered system, such as magnons, could be used to transport polaron-polaritons.

\begin{acknowledgments}
We thank H.-T. Lim, E. Togan and Y. Tsuchimoto for extensive help in the fabrication of the sample. We also acknowledge fruitful discussions with O. Cotlet, E. Demler, T. Ihn, P. M\"arki, A. Popert, R. Schmidt, Y. Shimazaki and T. Smolenski. This work was supported by the Swiss National Science Foundation (NCCR Quantum Science and Technology). This project has received funding from the European Research Council under the Grant Agreement No 671000 (POLTDES).  
\end{acknowledgments}

%

\clearpage

\begin{center}
  \textbf{\large Supplementary Material} \\
\end{center}
\setcounter{equation}{0}
\setcounter{figure}{0}
\setcounter{table}{0}
\setcounter{page}{1}
\setcounter{footnote}{0}
\renewcommand{\theequation}{S\arabic{equation}}
\renewcommand{\thefigure}{S\arabic{figure}}

\section{Optical Setup}
\label{A:Optical_Setup}
\begin{figure}[h]
\includegraphics{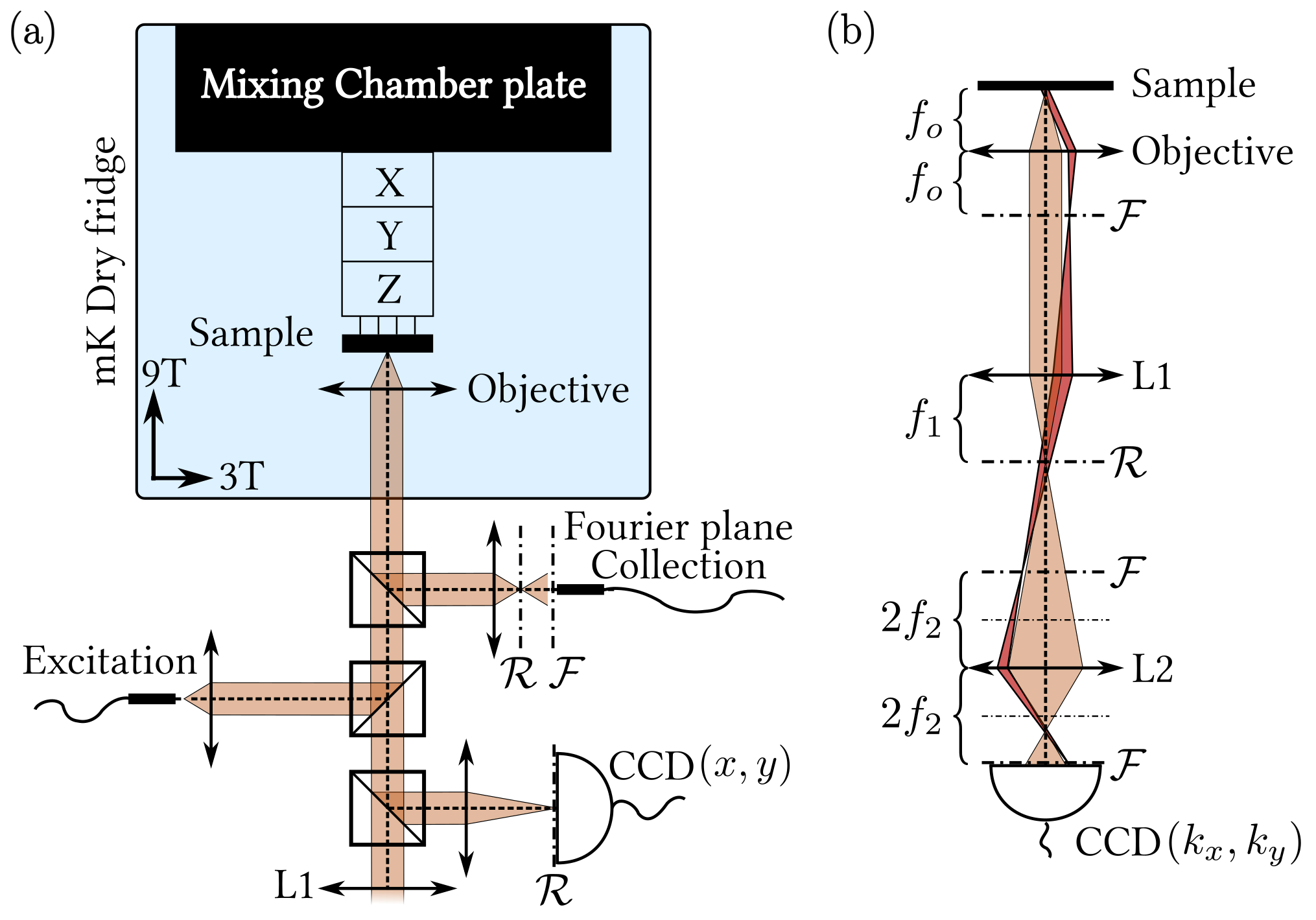}
\caption{\label{fig:Setup} Schematic of the optical setup. (a) Upper part of the cage assembly showing the excitation and collection arms and the real space imaging CCD camera. (b) Schematic ray tracing for Fourier space imaging. The intermediate real space image planes ($\mathcal{R}$) and Fourier space image planes ($\mathcal{F}$) are shown in black dashed lines.}
\end{figure}

Figure S1 depicts the optical setup used in this study. The free space optics used for excitation, collection and imaging are mounted in a cage assembly that is attached to the bottom of the dilution refrigerator (BlueFors LD250 with American Magnetics Inc. $9 \us \tx{T} + 3 \us \tx{T}$ vector magnet). Optical windows through the different shields of the cryostat offer free space optical access to the sample which is thermally anchored on the mixing chamber at a base temperature of $20 \us \tx{mK}$. The sample is electrically contacted as described in the main text and mounted on XYZ nanopositioners (Attocube ANPx101/RES, ANPz102/RES). A cold, high numerical aperture (NA) objective lens (Thorlabs 354330-B) allows for high resolution microscopy of the sample at base temperature. The excitation light (white light: Exalos EXS210036-01, resonant laser: Sacher TEC500 or PL laser: Melles Griot 05-SRP-812) is delivered by a single mode optical fiber and is collimated and directed to the microscope objective lens. The light emitted or reflected by the sample is refracted by the same objective lens and is simultaneously imaged onto real space and Fourier space CCD cameras (FLIR CM3-U3-13S2M-CS) as well as on the facet of a single mode optical fiber conjugated with the back-focal plane of the microscope objective. This collection fiber is mounted on an XY motrized stage (Thorlabs PIAK10) and is connected to a spectrometer (Princeton Instruments Acton SP2500), allowing to record spectra at different angles within the NA of the objective lens. Conversely, light can be sent via this collection fiber in order to excite polaritons with finite in-plane momentum. Due to the long distance between the sample and the bottom of the dilution refrigerator (\textit{ca.} $50 \us \tx{cm}$), the optical setup is not fully conjugated and relies instead on an image forming lens L1 and a 2f optical relay lens L2. The excitation, detection and imaging arms are all equipped with linear polarizers (Thorlabs LPVIS050) and quarter-wave plates (Thorlabs WPQ05M-808).

In order to account for the spectral shape of the white light source, the reflection spectra shown in the main text are corrected by the following procedure. First a numerical low-pass filter is applied to remove fast spectral etaloning fringes. The slowly varying spectral shape of the white light is then extracted using the fact that the polariton signal disperses as a function of some tuning parameter (position, momentum or magnetic field), while the lamp shape remains constant. By sorting the individual spectra along this tuning parameter axis we can extract the lamp shape and use it to normalize the reflection spectra.  

\section{Electrical transport properties}
\begin{figure}
\includegraphics{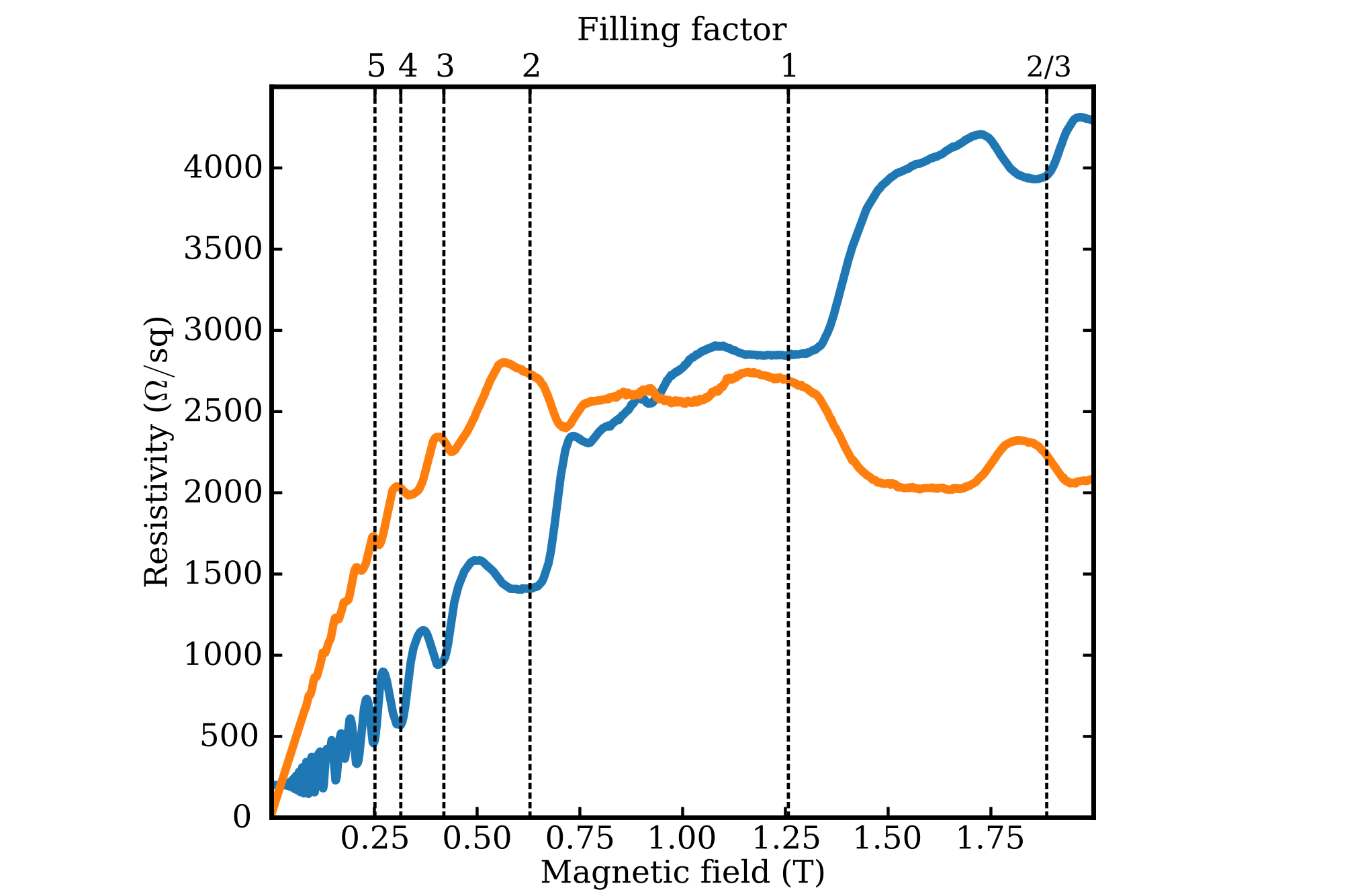}
\caption{\label{fig:Transport} Magneto-transport characterization. Longitudinal (blue) and transverse (orange) resistivities versus magnetic field and Landau level filling factor.}
\end{figure}

As shown in Fig. 1(b) of the main text, the annealing technique used to contact the 2DEG also contacts the low mobility doping QWs. Our device thus corresponds to a field effect transistor where the gate potential, nominally defined by the donor impurities, can be modified by the source-drain bias leading to the pinch-off of the 2DEG \cite{fritzsche_heterostructures_1987}. We show, in Fig.~S2, the longitudinal (blue curve) and transverse (orange curve) resistivities as a function of magnetic field, recorded using two lockin amplifiers (Signal Recovery 7265, $13.8 \us \tx{Hz}$ modulation, $1 \us \tx{nA}$). We clearly identify Shubnikov-de Haas oscillations in the longitudinal resistivity and the onset of Hall plateaus in the transverse resistivity. The existence of parallel conduction channels, in particular through the doping QWs, results in an overall trend in the resistivity curves that deviates from standard quantum Hall transport. This effect, well known in 2DEG transport \cite{kane_parallel_1985, reed_investigation_1986}, results from a trade-off in our device geometry between optimizing the optical properties and retaining good transport characteristics. It can be suppressed by thinning down the doping QWs, at the expense of an increased light-sensitivity of the device \cite{ravets_polaron_2018}.

\section{Optical Polarization Eigenbasis}
\begin{figure}
\includegraphics{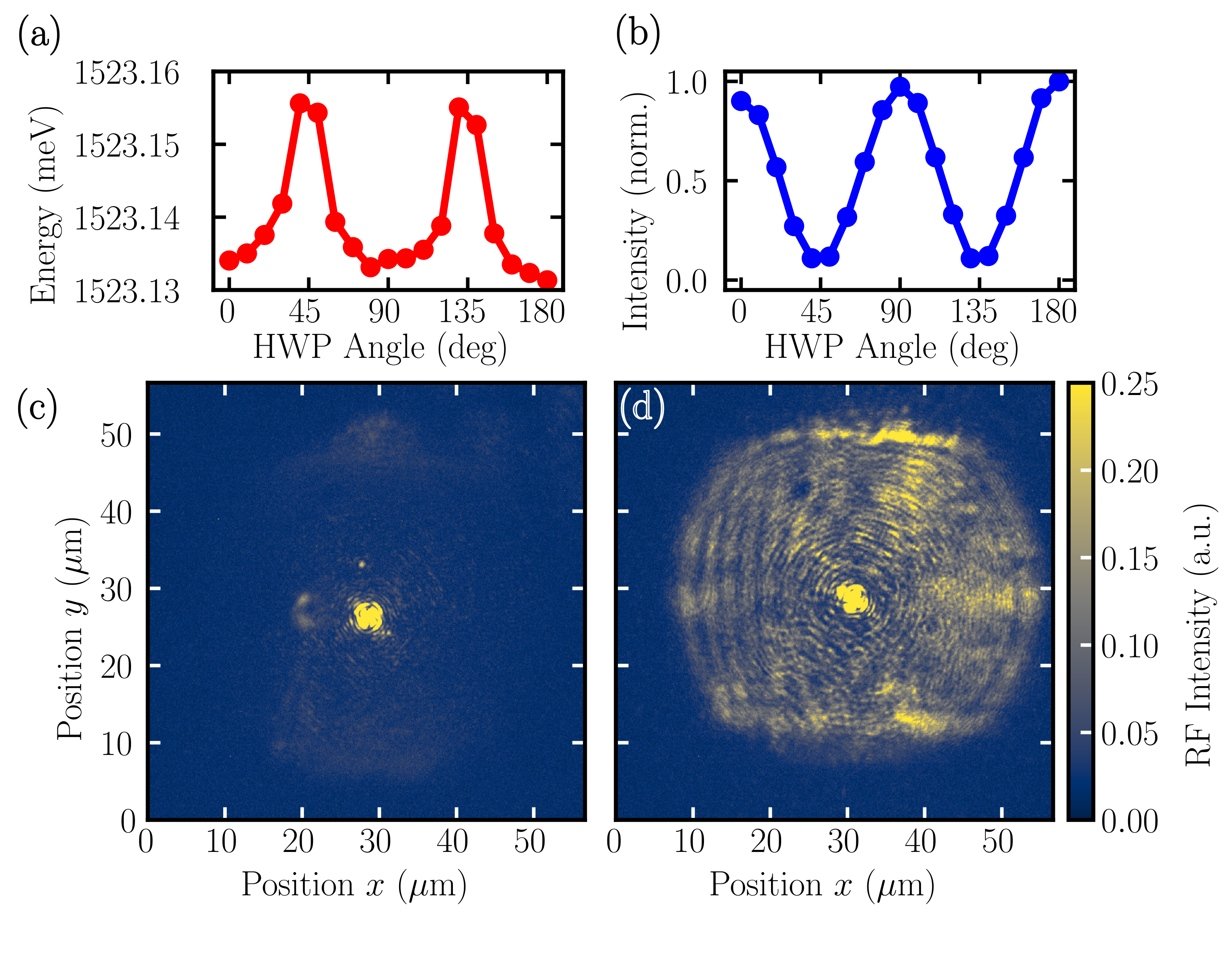}
\caption{\label{fig:polarization} Polarization eigenbasis defined by strain. Lower polariton emission energy (a) and intensity (b) as function of linear polarization angle. (c,d) Real-space images measured in cross-linear polarization at (c) $0^\circ$ and (d) $45^\circ$ with respect to the Hall bar $x$-axis.}
\end{figure}

In this section, we discuss the optical polarization properties of our sample and how they were used to obtain the measurements in resonance fluorescence (RF) configuration. In a first experiment, we excite the system non-resonantly at $632 \us \tx{nm}$ and measure the $k_\parallel = 0$ lower polariton photoluminescence center energy (Fig.~S3(a)) and intensity (Fig.~S3(b)) as a function of polarization angle. A half-wave plate in front of a horizontal polarizer was used to rotate the detection angle. We observe an energy splitting of $\Delta_{xy} = 20\us \tx{\mu eV}$ between the linear polarization eigenstates pointing along the crystalline axes of our sample. Note that these axes also align with the Hall bar and thereby the $x$- and $y$-axes used in all real-space images. A second experiment was performed with resonant excitation at finite $k_\parallel$ to verify that this birefringence still dominates compared to the TE-TM mode splitting at excitation angles relevant to our experiments. In Fig.~S3(c), polaritons are excited with polarization along $x$ and detected along $y$ which should be compared to Fig.~S3(d), where polaritons are excited at $45^\circ$ and detected at $-45^\circ$. The fact that the polariton cloud is still well suppressed in Fig.~S3(c) shows that the eigenbasis is aligned with the $x$, $y$ axes.

\section{Coupled Oscillator Model}
\begin{figure}
\includegraphics{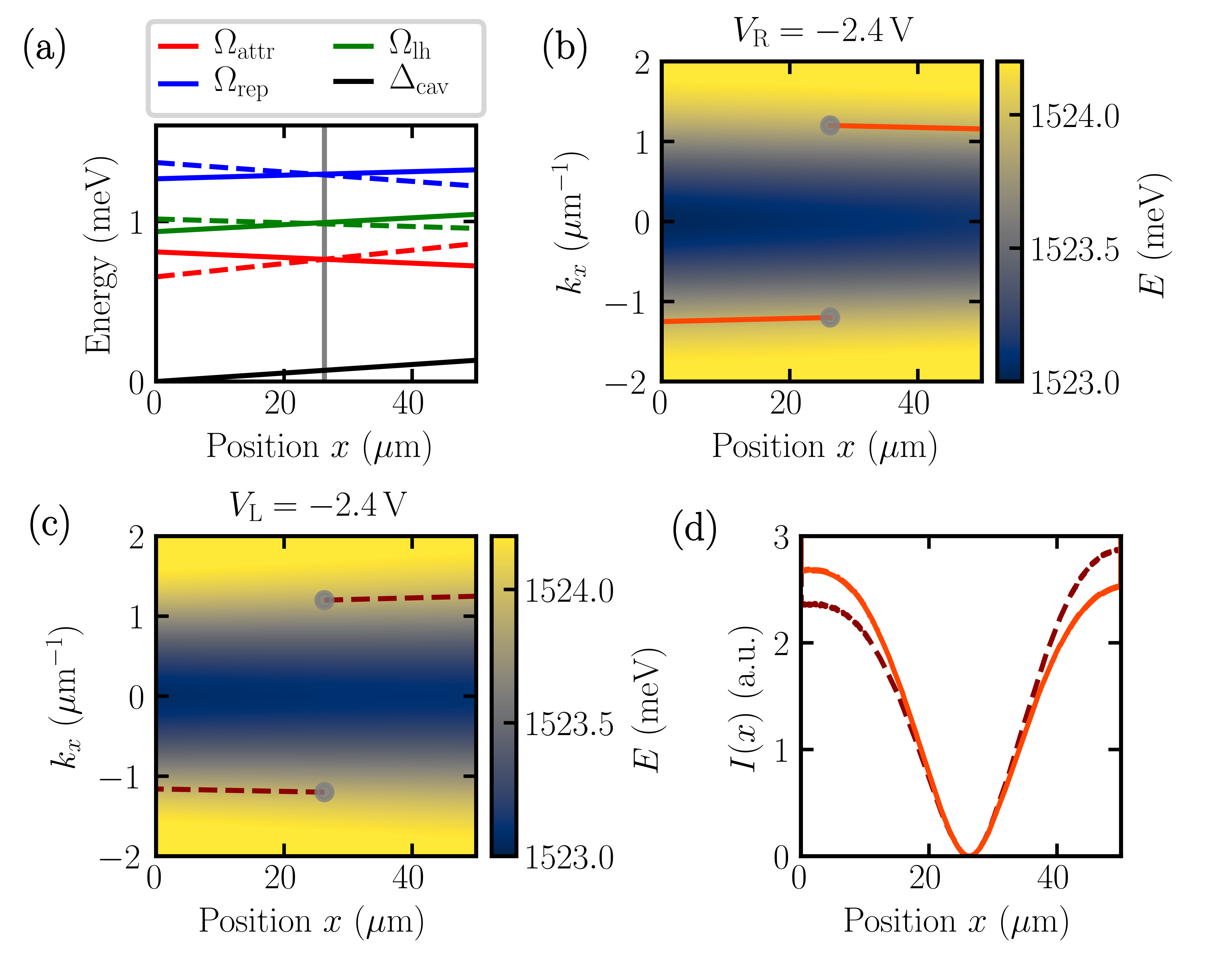}
\caption{\label{fig:coupled_oscillator} Coupled oscillator model to predict polariton acceleration. (a) Dispersion fit results. (b) Corresponding $E_\tx{LP}$ for depletion on the right and (c) depletion on the left side. Trajectories of lower polaritons are drawn with initial values marked by grey dots. (d) Expected emission from the trajectories shown in (b,c).}
\end{figure}

In the main text, we demonstrated the generation of electron density gradients (Fig.~2) and showed how they can be used to accelerate polaritons (Fig.~3(a)). Here, we characterize the energy $E_\tx{LP}(x, k_x)$ of the lower polariton branch and calculate the resulting forces for polaritons. This allows us to predict the changes in polariton group velocity and finally the expected shape of the polariton cloud subject to this force (Fig.~3(b)). We restrict ourselves to the one-dimensional case of propagation along the $x$-direction which should suffice for the experiments without magnetic field. We identify the cavity thickness variation and the electron density gradient as most important contributions to LP energy variations. Starting from the conditions in Fig.~3(a), we acquire four polariton dispersion relations: At the left side of the field of view ($x=0$) for both applied bias voltages ($\Delta V_\tx{R}= -2.4 \us \tx{V}$ and $\Delta V_\tx{L}= -2.4 \us \tx{V}$) and the same for the right side of the field of view ($x=50 \us \tx{\mu m}$). We fit the following model to each of the four dispersions
\begin{equation}
    \begin{pmatrix} 
        E_\tx{cav}(x, k_x)       & \Omega_\tx{attr}(x) & \Omega_\tx{rep}(x) & \Omega_\tx{lh}(x) \\
        \Omega_\tx{attr}(x)      & E_\tx{attr}         & 0                  & 0 \\
        \Omega_\tx{rep}(x)       & 0                   & E_\tx{rep}         & 0 \\
        \Omega_\tx{lh}(x)        & 0                   & 0                  & E_\tx{lh} \\
    \end{pmatrix}
\end{equation}
where we keep the energies of the asymptotes fixed, namely the attractive polaron energy $E_\tx{attr} = 1524.4 \us \tx{meV}$, the repulsive polaron energy $E_\tx{rep} = 1525.9 \us \tx{meV}$ and the light-hole $E_\tx{lh} = 1529.5 \us \tx{meV}$. This simplifies the fitting by reducing the number of free parameters and is justified by observing that the Rabi couplings vary more strongly than the energies as a function of electron density. The cavity dispersion was approximated as parabola $E_\tx{cav}(x, k_x) = E_0(x) + \hbar^2 k_x^2 / (2 m_\tx{cav})$ with cavity mass $m_\tx{cav} = E_0 {n_\tx{cav}^2}/{c^2} \approx 3\cdot 10^{-5} m_e$ measured independently. The cavity wedge in this region is $2.7 \us \tx{meV/mm}$ and the cavity detuning $\Delta_\tx{cav}$ is measured from $E_0 = 1524.2 \us \tx{meV}$ at $x=0$. The results are displayed in Fig.~S4(a) with full lines referring to bias applied to the right and dashed lines to bias applied to the left contact. The Rabi couplings and cavity wedge have been interpolated linearly between the two measured points at $x=0$ and $x=50\us\tx{\mu m}$. By diagonalizing Eq.~(S1), we extract the LP energy $E_\tx{LP}(x, k_x)$ for right and left bias, see Fig.~S4(b) and (c). Following the approach of \cite{steger_long-range_2013}, we write Hamilton's equations of motion for the lower polariton
\begin{equation}
    \frac{\partial x}{\partial t} = \frac{1}{\hbar} \frac{\partial E_\tx{LP}}{\partial k_x} \tx{\ and\ } 
    \frac{\partial k_x}{\partial t} = -\frac{1}{\hbar} \frac{\partial E_\tx{LP}}{\partial x},
\end{equation}
under which we propagate trajectories corresponding to the initial conditions of our experiment $x(t \mkern1.5mu{=}\mkern1.5mu 0) = 26\us \tx{\mu m}$ and $k_x(t \mkern1.5mu{=}\mkern1.5mu 0) = \pm 1.2 \us \tx{\mu m}^{-1}$. These trajectories are shown in Fig.~S4(b,c) for the two cases of bias voltages. In order to predict the RF signal obtained in our experiment we consider two more ingredients. First, the polaritons decay while propagating with a lifetime of $\tau \approx 10 \us \tx{ps}$. Second, to model the RF configuration we assume that the injected polarization state at $45^\circ$ slowly rotates in accordance with the polarization splitting determined in the previous section $\Delta_{xy}/\hbar \approx 0.03 \us \tx{ps}^{-1}$. Since we detect a real-space image, we create a histogram of the $x$-positions visited by the trajectories and weight them with $w=\exp{(-t/\tau)} \cdot \left( 1-\cos(\Delta_{xy} t / \hbar)\right)$ to obtain Fig.~S4(d). The normalized difference of this histogram is shown in the main text, Fig.~3(b). This analysis neglects variations of the effective mass (or polariton cavity content) during propagation. We verified that the result does not depend sensitively on the parameters $\tau$ and $\Delta_{xy}$. We want to point out that for the interpretation of the real-space images, one has to take into account the effective mass of the polaritons, which might be negative. Consider the full line in Fig.~S4(d), where polaritons are accelerated to the left. Due to the large in-plane momentum $k_x$, the acceleration actually reduces the group velocity during propagation. This leads to an excess of polaritons on the left side compared to acceleration in the opposite direction (dashed line). If we had access to a larger field of view, we would expect the normalized difference to change sign. The total number of polaritons stays the same when subjected to accelerating potentials but the position where they are re-emitted as photons changes.

\section{Electrically controlled spin density gradients}
\begin{figure}
\includegraphics{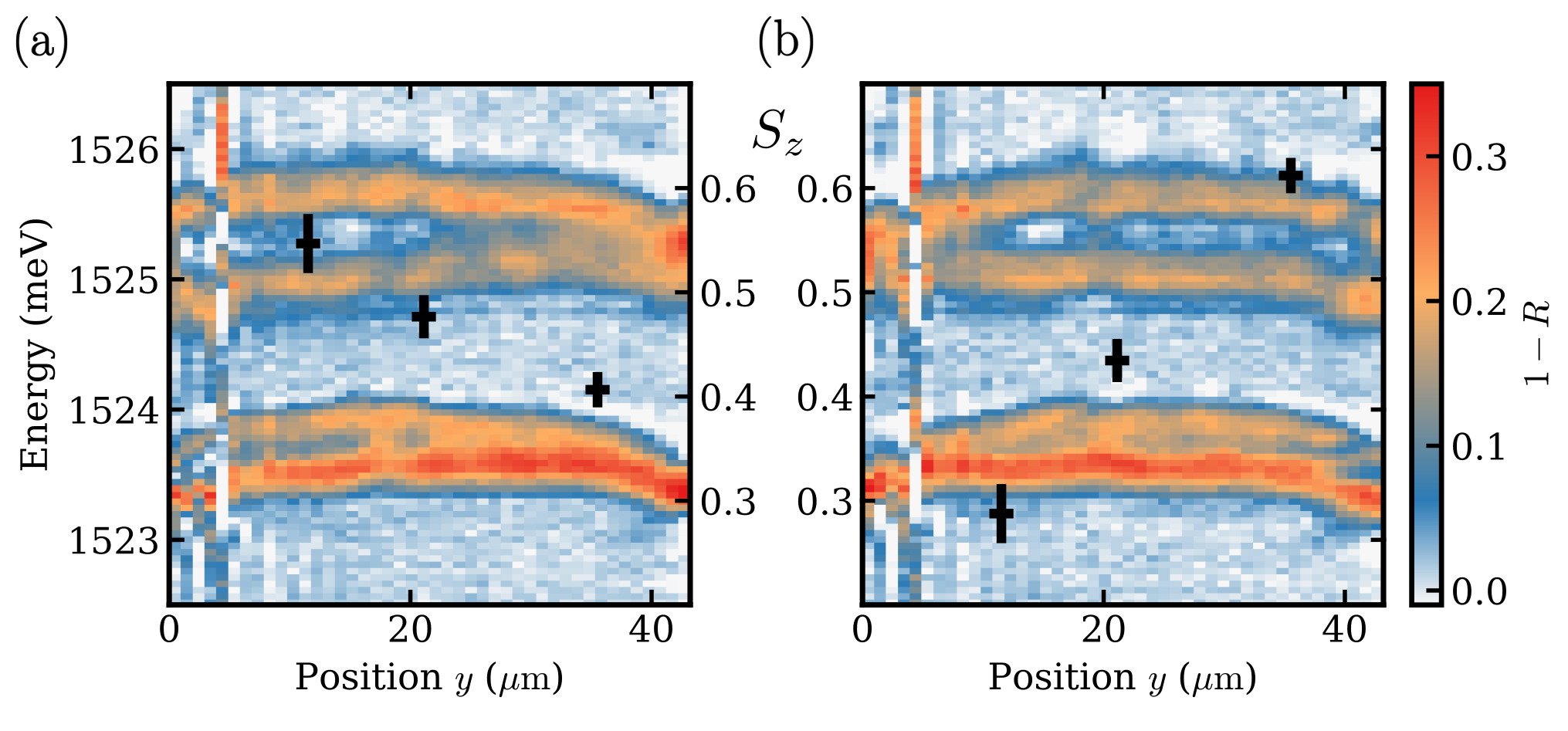}
\caption{\label{fig:SSH} Spatial tuning of polariton energy landscape in the quantum Hall regime. (a) Same as Fig.~4(b) of the main text and as comparison (b) with opposite spin density gradient.}
\end{figure}

We supplement Fig.~4 of the main text by showing that the spin polarization gradient around $\nu=1$ is tunable by the external bias similarly to what was shown for the electron density gradient. For this, we compare in Fig.~S5 two spatial maps of the polariton reflectivity taken at opposing voltages (a) $\Delta V_\tx{R} = -0.17 \us \tx{V}$ and (b) $\Delta V_\tx{L} = -0.35 \us \tx{V}$. These voltages correspond to a balanced situation where the filling factor at $y=25 \us \tx{\mu m}$ is equal and the gradient opposite. To assess the degree of spin polarization $S_z$ of the 2DEG, we measure polariton dispersions in $\sigma^+$ and $\sigma^-$ polarizations for the two opposing source-drain biases at three different $y$-positions. The coupled oscillators fit to each of these dispersion spectra yields values of the Rabi splittings entering the definition of $S_z$ at these three spatial positions and for the two opposing source-drain biases. The resulting $S_z$ values are reported as black dots in panels (a) and (b) with error bars extracted from the fit parameters.

\end{document}